\documentclass[amsmath,amssymb,aps,prl,reprint,floatfix,superscriptaddress]{revtex4-2}
\usepackage{placeins}

\usepackage{graphicx}
\usepackage{dcolumn}
\usepackage{xcolor}
\usepackage{bm}
\usepackage[version=4]{mhchem}
\usepackage[breaklinks,colorlinks,linkcolor={blue}, %
            citecolor={blue},urlcolor={blue}]{hyperref}
\usepackage[normalem]{ulem} 
\begin{document}

\title{Identifying high performance spectrally-stable quantum defects in diamond}
\author{Yihuang Xiong} 
\affiliation{Thayer School of Engineering, Dartmouth College, Hanover, New Hampshire 03755, USA}
\author{Yizhi Zhu} 
\affiliation{Thayer School of Engineering, Dartmouth College, Hanover, New Hampshire 03755, USA}
\author{Shay McBride}
\affiliation{Thayer School of Engineering, Dartmouth College, Hanover, New Hampshire 03755, USA}
\author{Sin\'ead M.\ Griffin}
\affiliation{Materials Sciences Division, Lawrence Berkeley National Laboratory, Berkeley, California 94720, USA}
\affiliation{Molecular Foundry Division, Lawrence Berkeley National Laboratory, Berkeley, California 94720, USA}

\author{Geoffroy Hautier} 
\affiliation{Thayer School of Engineering, Dartmouth College, Hanover, New Hampshire 03755, USA}
\date{\today}

\begin{abstract}
Point defects in semiconductors are becoming central to quantum technologies. They can be used as spin qubits interfacing with photons, which are fundamental for building quantum networks. Currently, the most prominent quantum defect in diamond is the nitrogen-vacancy (NV) center. However, it suffers from spectral diffusion that negatively impacts optical coherence and is due to the coupling of the emission energy with uncontrolled electric fields. The group IV vacancy complexes on the other hand have shown to be significantly more spectrally-stable as they are centrosymmetric and thus immune to the linear Stark shift. They however suffer from several issues ranging from low operation temperature to low optical efficiency due to dark states and difficulty in stabilizing the right defect charge state. Here we search for alternative to the group IV vacancy complex in diamond by systematically evaluating all possible vacancy complex using high-throughput first-principles computational screening. We identify the defects that combine centrosymmetry, emission in the visible range, as well as favorable and achievable electronic structure promoting higher operation temperature and defect levels well within the band gap. We identify: for which we find ZnV2- to be especially appealing. 

\end{abstract}

\maketitle

\section{Introduction}

Quantum defects are color centers in semiconductors used for quantum applications (sensing or communication). At low concentrations, these defects remain isolated from environmental perturbations due to the host material, yet they can still be initialized and read out using electromagnetic fields\cite{Bassett_nanoph_2019, Wolfowicz2021, Awschalom_nature_photonics_2018}. These properties make quantum defects appealing platforms for developing spin-photon interfaces, which serve as fundamental building blocks for quantum communication networks via flying photons\cite{Simmons.PRXQuantum.2024, Stolk_PRX_2022,Komza_nat_comm_2024}. Currently, the most prominent quantum defect is the nitrogen-vacancy (NV) center in diamond. When negatively charged, the NV center possesses a triplet ground state and emits brightly with a nanosecond radiative lifetime. The NV center has shown promising progress in quantum metrology and communication \cite{Atature2018, Awschalom_nature_photonics_2018, Pompili_science_2021, Katsumi_Comm_Mater_2025}. Nevertheless, the NV center is still far from perfect. One of its main drawbacks is that its zero-phonon line (ZPL) emission is strongly perturbed by fluctuations in the electric field, which may be due to the presence of nearby defects or surface states. Such variations in the ZPL, known as spectral diffusion, limits optical coherence as it deteriorates the indistinguishability of the emitted photons and compromise its performance for entanglement generation.\cite{Orphal-Kobin_PRX_2023}.
\begin{figure}[!ht]
 	\centering
 	\includegraphics[width=0.5\textwidth]{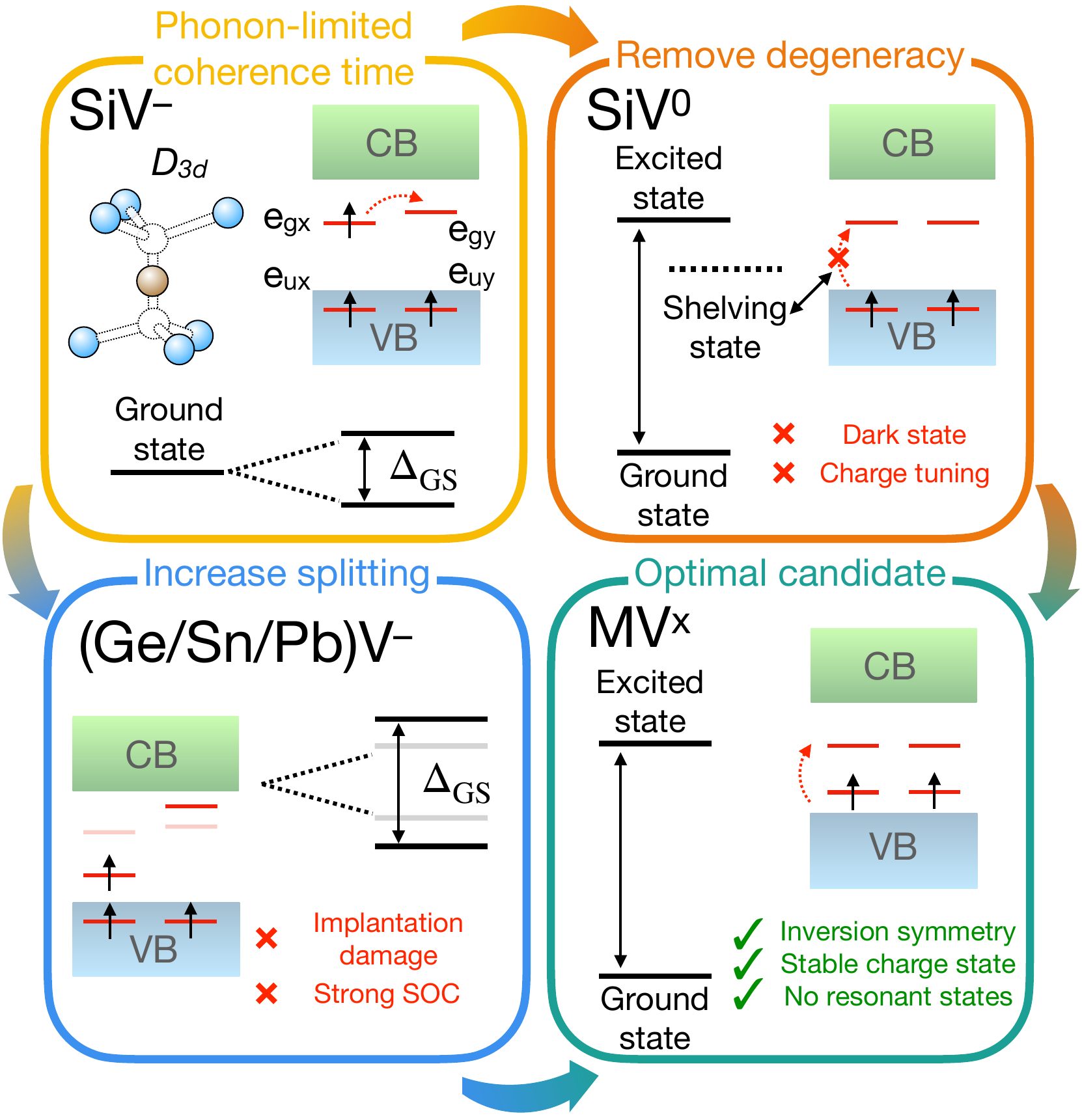}    
 	\caption{The main drawback of Si$V^-$ the phonon-limited coherence time. Current strategies to mitigate this are 1) through charge tuning to remove degeneracy (Si$V^0$). Fermi level control is difficult, and also cause dark state; 2) apply strain or chemical strain (heavier atoms) to further split the energy levels, however, could results in higher implantation damage and shorter spin coherence due to spin-orbit coupling.
 	}
 	\label{Fig.1}
\end{figure}

The ZPL emission sensitivity to static electric fields is due to the different in Stark shift between the ground and excited state. Although methods have been developed to mitigate the Stark shift through surface termination\cite{Sangtawesin_PRX_2019}, nanostrucutre \cite{Ruf_Nano_letters_2019, Orphal-Kobin_PRX_2023}, or active control schemes \cite{Fotso_PRL_2016}, the suppression is still insufficient. An alternative approach is to use quantum defects that are intrinsically immune from Stark effect when the defect structure as an inversion symmetry\cite{Sipahigil_PRL_2014,Brendon_science_2018}. This development of spectrally-stable quantum defects through symmetry constrained has been exemplified by the group IV vacancy complex such as Si$V$ and Sn$V$. These defects form in a splitting-vacancy configuration with a $D_{3d}$ point group (see Figure \ref{Fig.1}). As expected, the group IV vacancy complexes have shown low spectral diffusion and high optical coherence \cite{Palyanov_Sci_Rep_2015, Sipahigil_PRL_2014, Iwasaki_PRL_2017,Brendon_science_2018,Gergo_2018_PRX}. In fact, the recent demonstrations of entanglement over large distances were performed using the silicon vacancy complex\cite{Knaut_Nat_2024}.

However, despite the high optical coherence of these vacancy complexes, they have several other issues. The first of these defects to be studied the negatively charged silicon vacancy complex (Si$V^-$) shows a short spin coherence time due to the phonon-mediated orbital relaxation\cite{Rogers_PRL_2014, Pingault_Nat_comm_2017} (see Figure~\ref{Fig.1}). This has been related to its electronic structure with egx and egy occupied and unoccupied orbitals very close in energy leading to a small split of orbital branches of the ground state ($\Delta$GS$\approx$47GHz)\cite{Becker_Nat_Comm_2016}. Spin coherence can be increased by lowering temperature and it is imperative to operate Si$V^-$ in millikelvin temperature to suppress phonon-mediated dephasing\cite{Pingault_PRL_2014}. 

To overcome this difficulty and being able to operate at higher temperature, it has been suggested to instead use the neutral silicon vacancy complex (Si$V^0$). As the $e_g$ states are not occupied anymore, the phonon-assisted decoherence process is suppressed (see Figure~\ref{Fig.1}).\cite{Brendon_science_2018} Unfortunately, Si$V^0$ is challenging to stabilize as it requires a Fermi level that is only achievable in highly boron-doped diamond or through specific band bending through surface termination\cite{Brendon_science_2018,Green_PRB_2019, Zhang_PRL_2023}. Additionally, though shelving states have also been suggested for Si$V^-$ in diamond \cite{Gali.PRB.2013, Neu_PRB_2012}, it is much more pronounced in its neutral charges, results in weak fluorescence at higher temperature and limiting the optical efficiency, due to the resonant defects states in the valence bands \cite{Green_PRB_2019, Garcia_Adv_Sci_2024}. An alternative approach has been proposed by using heavier group IV elements (Ge, Sn, Pb), the $\Delta$GS can be increased by creating "chemical" strain\cite{Gergo_2018_PRX, Bradac_Nat_Comm_2019} (see Figure~\ref{Fig.1}). Such effect is limited for Ge$V^-$, and can introduce other spin-dephasing mechanism such as strong spin-orbit coupling. Heavier elements could also cause larger lattice damage that impact the performance. Lower yield has been observed for heavier elements after implantation\cite{Bradac_Nat_Comm_2019}. It is also noted that beside Si$V^0$, no other neutral defects have been clearly identified with optical emission in experiments. It has been reported that resonant defects states still present for these color centers, and could lead to dark states and photoionization instability \cite{Ma_pccp_2020, Gergo_2018_PRX}. Among the group IV elements beyond silicon, the Sn vacancy complex has been the most used and studied.

In this work, we aim to systematically search for all potential vacancy complex in diamond using first principles high-throughput computations. We aim at finding a centrosymmetric defect with an optimal electronic structure that would be stable within the typical Fermi level ranges in diamond and present no resonant states from the VBM (see Figure ~\ref{Fig.1}). We screened more than 500 substitution-vacancy complex (M$V$) defects in diamond, and further identified a series of promising defects statisfying those constraints. Among these defects, we further study the Zn$V$ complex in the $-$2 charge state. It has a triplet ground state and emits brightly in the visible range. Zn$V^{-2}$ possesses a similar electronic structure to neutral Si$V^0$, however, without the resonant states as the defect levels are well isolated from the band edge, avoiding potential dark states due to the valence bands. By investigating its zero-field splitting and vibronic properties, we confirm that  Zn$V^{-2}$ is a promising quantum defects. Lastly, we analyzed the defect database and unveil the chemical rules of designing quantum defects without stark shift while maintaining appealing optical properties. 

\section{Results}

\subsection{First-principles high-throughput screening}

\begin{figure}[!ht]
 	\centering
 	\includegraphics[width=0.45\textwidth]{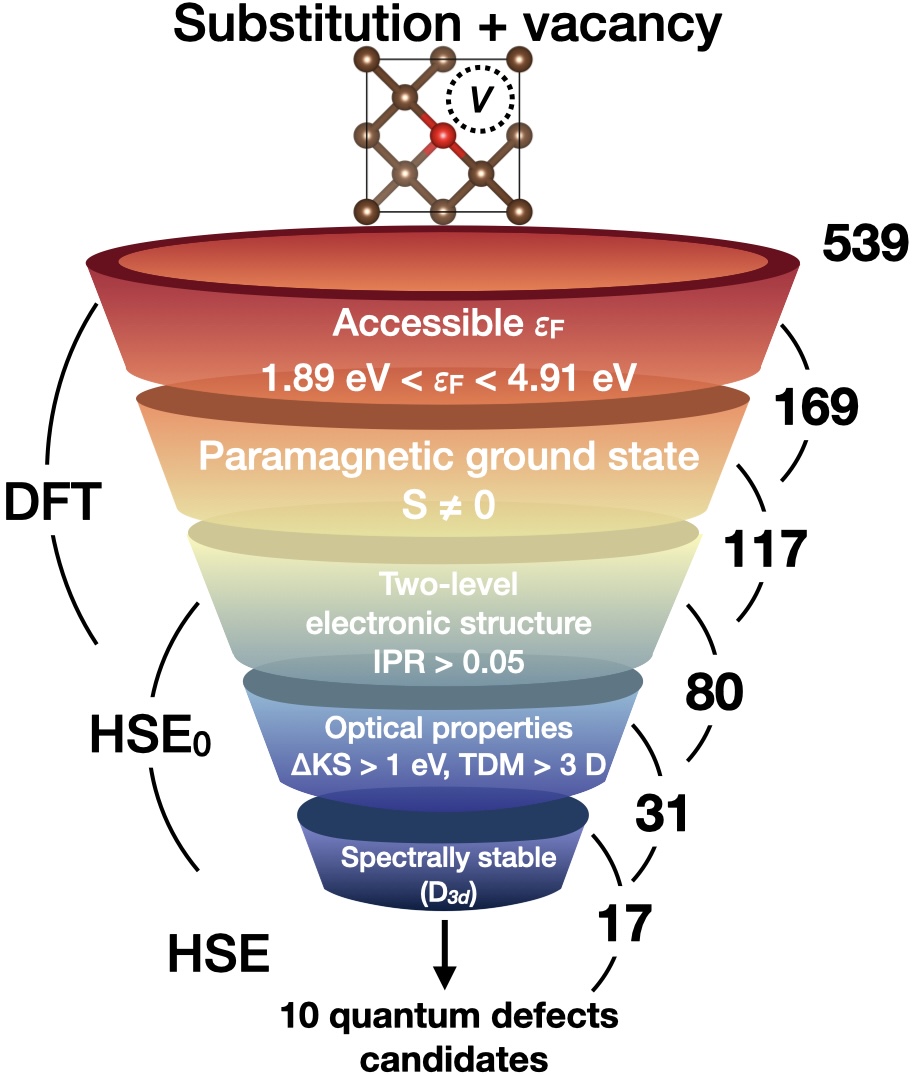}    
 	\caption{High-throughput screening workflow of spectrally stable quantum defects focusing on substitution-vacancy complex defects with D$_{3d}$ symmetry.
 	}
 	\label{Fig.2}
\end{figure}

\begin{figure}[t]
 	\centering
 	\includegraphics[width=0.45\textwidth]{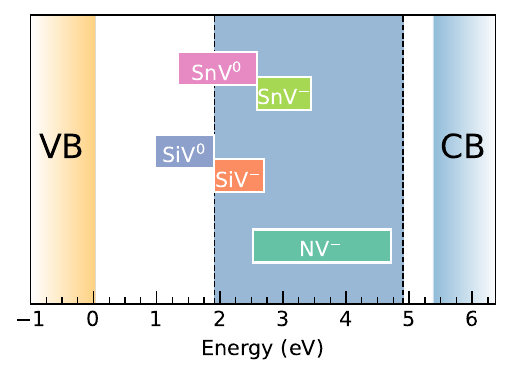}    
 	\caption{synthesizability figure
 	}
 	\label{fermi_synthesizability}
\end{figure}
We began by constructing a defect database encompassing the majority of the periodic table, excluding rare-earth elements and noble gases. Hydrogen was also excluded due to complexities arising from its small atomic radius affecting defect structures. Motivated by the inversion symmetry observed in group IV-vacancy defects, we primarily focus on the substitution-vacancy (MV) complex defects in the database, totaling 539 charged defects. Each complex defect structure was initialized in a lower symmetry C$_{3v}$ configuration (same as NV center) to avoid structure trapping at high-symmetry local minimum. Following structural optimization with no symmetry constraint at semilocal DFT (GGA-PBE) level, we employed a single-shot hybrid functional (HSE$_0$) calculation to mitigate the orbital energy underestimation associated with semilocal functionals. This approach has been demonstrated to enhance the accuracy of electronic structure descriptions at minimal additional computational cost~\cite{Alkauskas_PhysicaB_2007,Yang_JPCL_2022,Thomas_Nat_comm_2024,Xiong2023,Xiong_jacs_2024}. See Methods section for further details.

We designed the screening criteria as illustrated in Figure~\ref{Fig.2}. One of the primary challenges in realizing quantum defects in diamond is the limited control over the Fermi level positions. Unlike silicon, diamond exhibits strong $sp^3$ bonding, which complicates doping efforts. Defects charge transition levels are often deep within diamond's wide band gap and has high activation energies\cite{KALISH1999781, CRAWFORD2021100613,Stenger_JAP_2013}. Currently, p- and n-type doping can be achieved using boron and nitrogen/phosphorus, respectively; however, the achievable Fermi level range is still limited compared to what can be achieved in silicon \cite{Mat_Sci_in_Semi_Proc_2025}.

\subsection{Screened candidates}
\begin{figure*}[!th]
 	\centering
 	\includegraphics[width=1\textwidth]{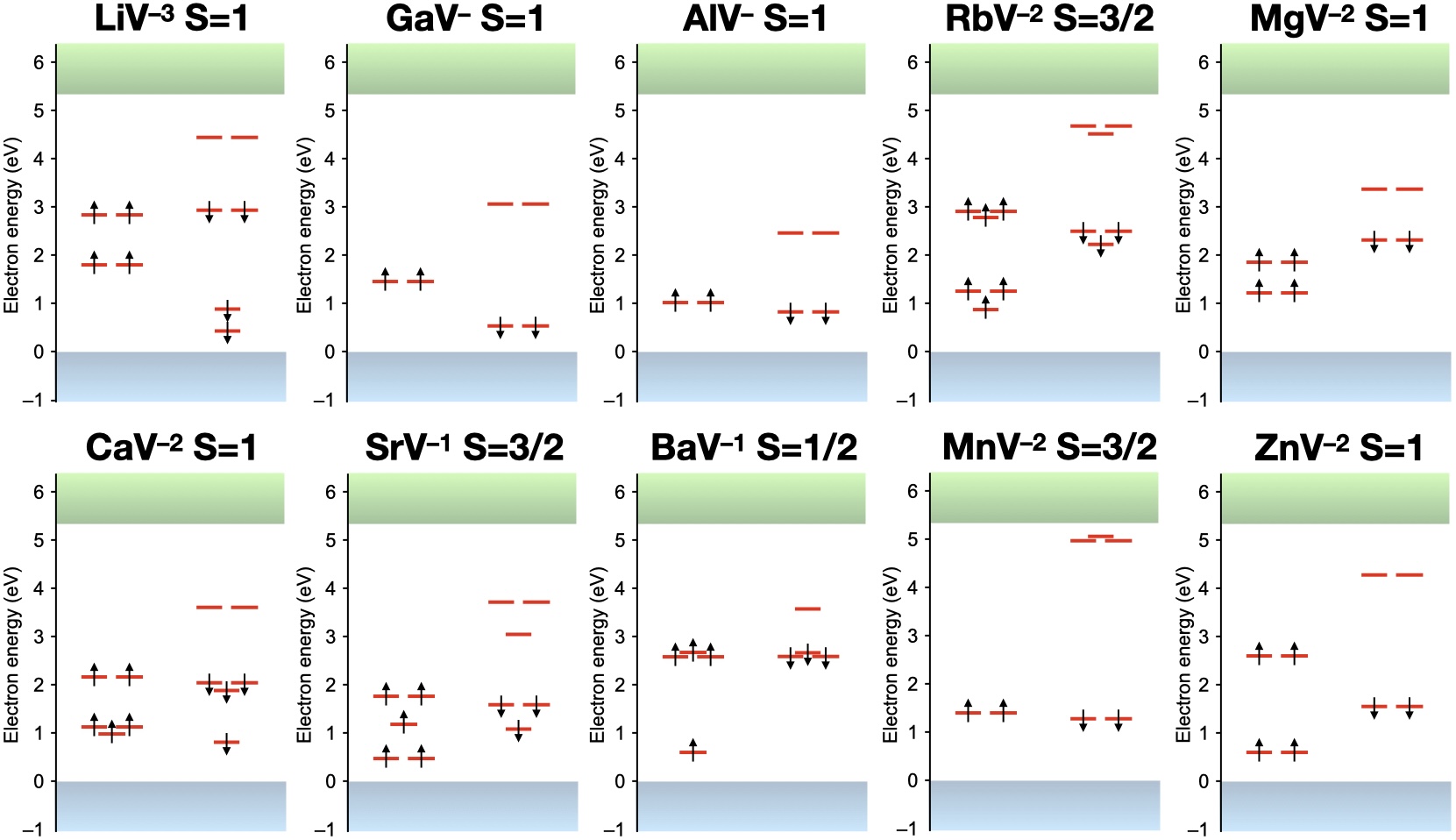}    
 	\caption{Single particle Kohn-Sham level computed at HSE level for the screened candidates. The charge state and the spin multiplicity are also labeled for each defect.
 	}
 	\label{screened candidates}
\end{figure*}

To ensure the practical accessibility of charged defects in diamond, we estimate the achievable Fermi level in diamond as plotted in Figure~\ref{fermi_synthesizability}. This range was determined based on specific considerations. The lower bound was set at the computed (0/$-1$) charge transition level for Si$V$, due to the extra synthesis requirements of B-doping or surface hydrogen passivation \cite{Brendon_science_2018,Zhang_PRL_2023}. The upper bound corresponds to nitrogen doping, which is always present in diamond samples. The high commercially available doping level for diamond optics of 10 ppm (Element Six DNV-B14) corresponds to a concentration of  1e18 cm$^{-3}$. Assuming complete ionization of nitrogen, we performed a simple estimation based on the Fermi–Dirac distribution and determined that annealing the sample at 1000 K sets the Fermi level at approximately 0.45 eV below the CBM (see Supplementary Note 1 for details). Consequently, the accessible range of the Fermi level in diamond spans from 1.91 to 4.91 eV relative to the valence band maximum VBM.

We overlay in Figure~\ref{fermi_synthesizability} the formation energies of the well-known N$V$ and Si$V$ systems. It is evident that N$V^-$ predominantly resides within the defined doping range, consistent with its ease of synthesis~\cite{weber_pnas_2010}. We also plot Sn$V^-$ and Sn$V^0$, showing their charge states are within the achievable range. Charge tuning experiments have indicated the existence of Sn$V^0$\cite{Luhmann_ACS_photonics_2020,Gorlitz_npj_QI_2022}. We note that the lack of optical characterization could be due to the weak photoionization stability\cite{Qiu_PRB_2023,Gergo_2018_PRX}. These could serve as additional considerations for thermodynamically stable defects.

Following the charge stability criteria outlined previously, we select defects with paramagnetic ground states (S$\neq$0) for spin-photon interface applications. Subsequently, we consider defects exhibiting two in-gap defect levels analogous to NV centers, which may enhance the operating temperature and brightness~\cite{Xiong.mqt.2024}. We note that bound-excitonic defects may still prove useful or even essential in hosts with smaller band gaps, such as T center in silicon~\cite{Dhaliah2022, Xiong2023, Xiong_jacs_2024}. To verify the presence of in-gap defect levels, we quantify the localization of Kohn-Sham (KS) orbitals with inverse participation ratio greater than 0.5 (further details in the Supplementary Information). As previously discussed, we employed the HSE$_0$ functional to achieve improved electronic structure descriptions~\cite{Xiong2023}. Utilizing these enhanced electronic structures, we approximate the zero-phonon line (ZPL) through KS energy differences ($\Delta$KS). Given that $\Delta$KS lacks of ionic relaxation and electron-hole interactions compared to ZPL, we set a threshold of $\Delta$KS \textgreater 1 eV to ensure sufficient emission wavelengths. By focusing on candidates with bright emission (transition dipole moment\textgreater3 D), we specifically screen defects possessing D$_{3d}$ symmetry, which exhibit vanishing electric dipole moments during optical excitation. Further methodological details are provided in the Methods section.

It is noteworthy that previous screening studies for color centers in diamond were conducted by Davidson and colleagues. Our work distinguishes itself in several key aspects: 1) addressing the critical need for spectrally stable color centers; 2) employing the HSE$_0$ for improved electronic structures that reveals candidates previously overlooked; and 3) targeting charged defects that are realistically synthesizable taking into account diamond's limited dopability. Based on this aforementioned screening, we identify 17 promising candidates as potential high-performance quantum defects. Their HSE$_0$ electronic structures are presented in Figure SX in the Supplementary Information.

We now focus on the identified candidates by refining their electronic structures using full HSE calculations in 512-atom supercells. These refined calculations reveals 10 out of the original 17 defects as promising candidates, and their corresponding HSE defect levels are displayed in Figure~\ref{screened candidates}. The candidate list was narrowed due to symmetry reductions at the HSE level. DFT often introduces partial occupancy in degenerate Kohn-Sham levels, artificially suppressing Jahn-Teller distortions. Hybrid functionals penalize partial occupancy, thus restoring Jahn-Teller distortions and resulting in reduced symmetry. We carefully verified the 10 final candidates to confirm their inversion symmetry. The computed $\Delta$KS values and transition dipole moments (TDM) evaluated using HSE wavefunctions are summarized in Table S1 of the Supplementary Information.

We note that several of the screened defects have been previously studied. For instance, the MgV complex defects were experimentally observed with a ZPL near 2.2 eV \cite{Luhmann_jpd_2018,Luhmann_nat_comm_2019}. Subsequent theoretical work by Pershin \textit{et al.} suggested that the third excitation of MgV$^-$ accounts for this ZPL, whereas the centrosymmetric MgV$^{2-}$ is unfortunately limited by photoionization \cite{Pershin_npj_qi_2021}. Our screening also highlights negatively charged group III vacancy complexes (GaV$^-$ and AlV$^-$), previously reported by Harris \textit{et al.} \cite{Harris.prb.2020}. Chemically, negatively charged group III vacancies mimic neutral group IV elements, with both $e_u$ and $e_g$ defect levels situated within the band gap, leading to favorable electronic structures. The identification of previously known defects underscores the robustness of our screening methodology. We also found that some well-known centrosymmetric quantum defects, such as SiV and NiV$^0$, do not appear in our final candidate list, as their defect levels resonate with the bulk bands, failing our criterion of having two distinct in-gap defect levels \cite{Gali.PRB.2013,Thiering.prr.2021}.

\begin{figure}[t]
 	\centering
 	\includegraphics[width=0.5\textwidth]{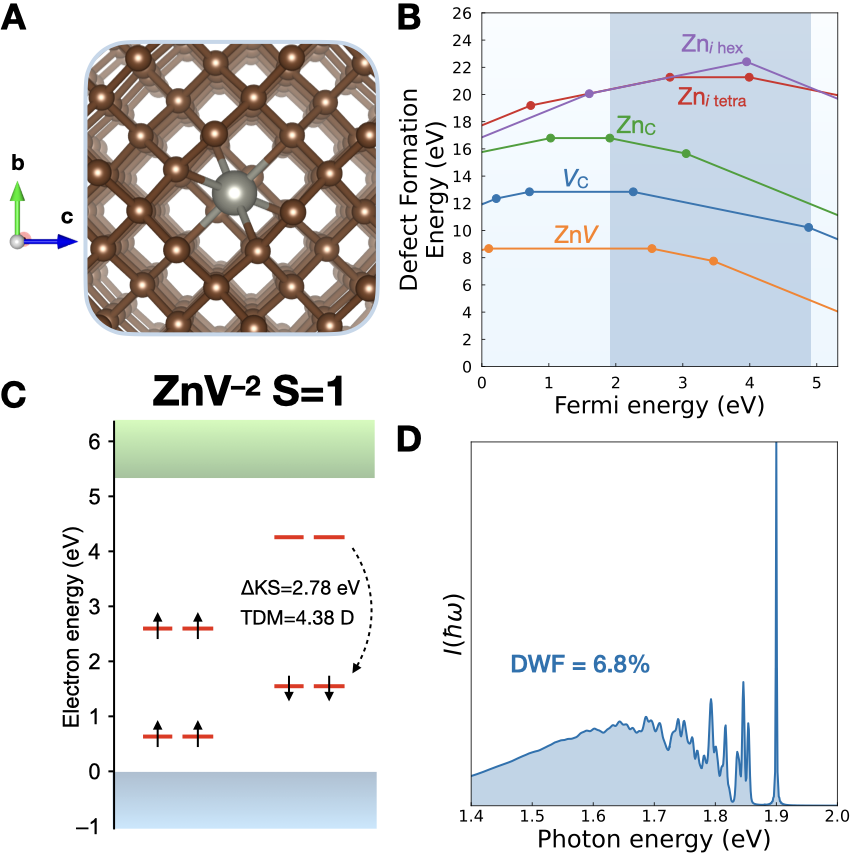}    
 	\caption{A.Defect structure of Zn$V^{2-}$ in $D_{3d}$ configurations. B. Defect formation energies Zn$V$ and its simple defect counterpart $V_{\rm C}$ and Zn$_{\rm C}$, the chemical potentials are referenced to the elemental form. C. Kohn-sham defect levels at HSE level predicts no defects states are resonant with the bulk bands. D. Simulated photoluminescence with and without Jahn-teller effect.
    }
 	\label{ZnV2-}
\end{figure}

For all the screened candidates, Ga$V^-$, Ca$V^-$, Al$V^-$, and Zn$V^{2-}$ have shown $\Delta$KS at HSE level of more than 1 eV. Since $\Delta$KS approximation does not taking into account of the relaxation, we anticipate the ZPL of these candidates will be in technologically relevant range. In combination with their inversion symmetry, making them appealing candidates for quantum applications. Among these candidates, we pay special attention to Zn$V^{2-}$. It possesses a triplet ground state and its electronic structure resembles Si$V^0$ with defect levels isolated from bulk bands, minimizing interference from valence-band-related dark states. The atomic and electronic structures of Zn$V^{2-}$ are shown in Figures~\ref{ZnV2-}(a) and (c). We computed the defect formation energies for Zn$V$ as well as its constituent simple defects (V$_\mathrm{C}$ and Zn$_\mathrm{C}$) at HSE level, showing in Figure~\ref{ZnV2-}(b). The charge stability region of Zn$V^{2-}$ covers the majority of the accessible Fermi level range, highlighting the accessibility of its charge state. Zn$V$ complexes exhibit lower formation energies compared to V$_\mathrm{C}$ and Zn$_\mathrm{C}$, showing a positive defect binding energy and a thermodynamic preference for its formation. We also note that Zn$V^{-}$ also resides within this stability region. With one fewer electron, Zn$V^{-}$ undergoes a symmetry reduction to C$_{2h}$ due to the JT effect. Additionally, an extra unoccupied $e_g$ state in the spin-majority channel introduces undesirable competing excitations, detailed in the Supplementary Information. Nevertheless, the concentration of Zn$V^{2-}$ might be readily enhanced through nitrogen doping, commonly available in diamond. We also want to highlight that the most of Zn isotopes are spin-free besides $^{67}$Zn. Since other forms of Zn-related defects might unavoidably present in diamond such as Zn$\rm _C$, this is beneficial for the nuclear spin contrast to group III-vacancy complex or P1 (N$\rm _C$) center in diamond. 

The electronic structure of Zn$V^{2-}$ is presented in Figure~\ref{ZnV2-}(c). The $e_u$ and $e_g$ states in the spin-minority channel are separated by 2.78 eV. The transition dipole moment calculated from HSE ground-state wavefunctions is 4.38 D, which is comparable to the other centrosymmetric defect such as SiV$^0$ (4.65 D), SiV$^-$ (6.06 D), and SnV$^-$ (5.36 D). The corresponding radiative lifetime is estimated to be 6.7 ns, indicative of bright emission. To further explore the optical properties of Zn$V^{2-}$, constrained HSE calculations were performed by promoting an electron from the $e_u$ to the $e_g$ state. The zero-phonon line (ZPL) energy corresponds to the energy difference between the excited and ground states with their atomic structure optimized. Upon removing symmetry constraints, the excited state of Zn$V^{2-}$ relaxed to $C_{2h}$ symmetry due to Jahn-Teller distortion, yielding a ZPL at 1.88 eV—approximately 60 meV lower than that of the NV center. We further simulate the photoluminance spectrum of Zn$V^{-2}$ under the the Huang-Rhys approximations by assuming the ground-state and excited-state phonon are identical \cite{Alkauskas_2014,huang1950theory}. Given the Jahn-Teller instability of the excited state, we simulated the photoluminescence lineshape based on the symmetry-reduced excited states, as shown shown in Figure~\ref{ZnV2-}(d). The computed Debye-Waller factor is around 6.8\%, comparable to that of the NV$^-$ center (approximately 3\%). We expect that the ZPL can be enhanced via optical cavities through the Purcell effect. This has been demonstrated for NV centers that the DWF can be improved up to 44\% \cite{Riedel_PRB_2017,Yurgens_2024_npj_QI}. Nevertheless Zn$V^{2-}$ exhibits a vanishing electric dipole moment and bright emission at 1.88 eV, combined with thermodynamic stability and easily accessible charge states, making it an attractive candidate for spin-photon interfaces in diamond.

\subsection{Chemical rules for designing high-performance quantum defects}

\begin{figure*}[t]
 	\centering
 	\includegraphics[width=1\textwidth]{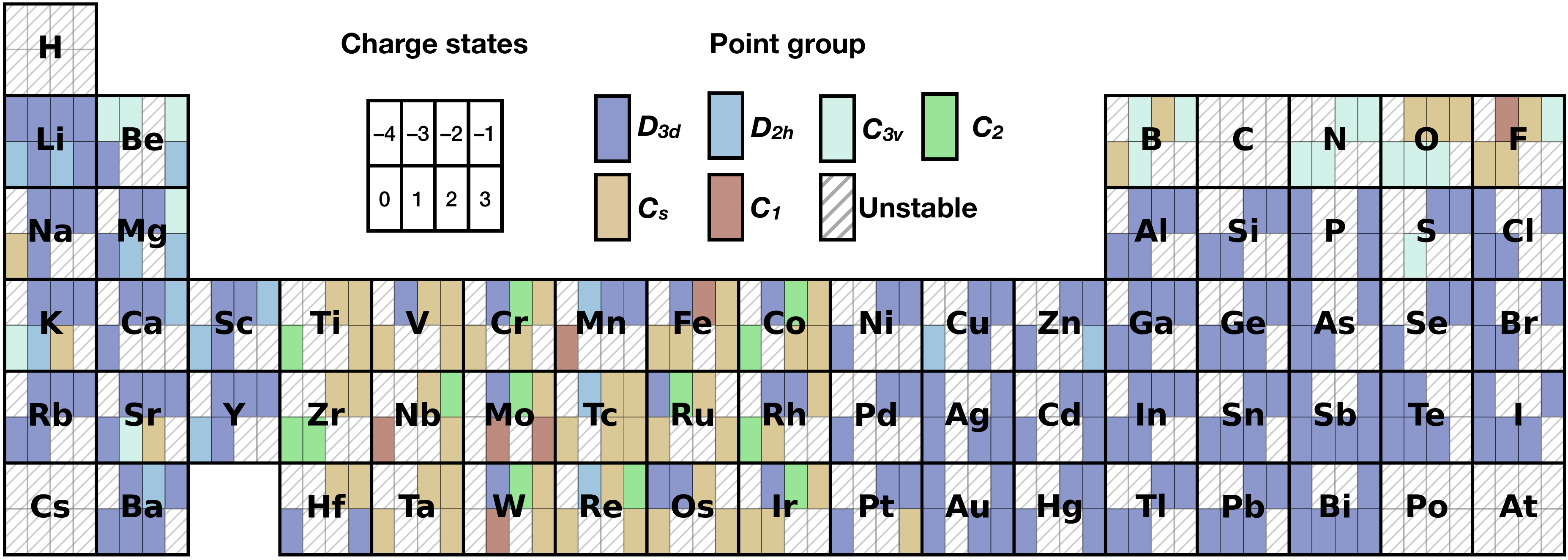}    
 	\caption{Charge stability and defect symmetry analysis for MV defect. Each elemental block is composed of eight mini-panels that stands for the charge states from -4 to 3, respectively. The corresponding defect point group is illustrated by the color codes. If the charged defect is not stable, a shaded grid is used to represent the corresponding state.
 	}
 	\label{Fig.periodic_table_trend}
\end{figure*}

\begin{figure}[t]
 	\centering
 	\includegraphics[width=0.5\textwidth]{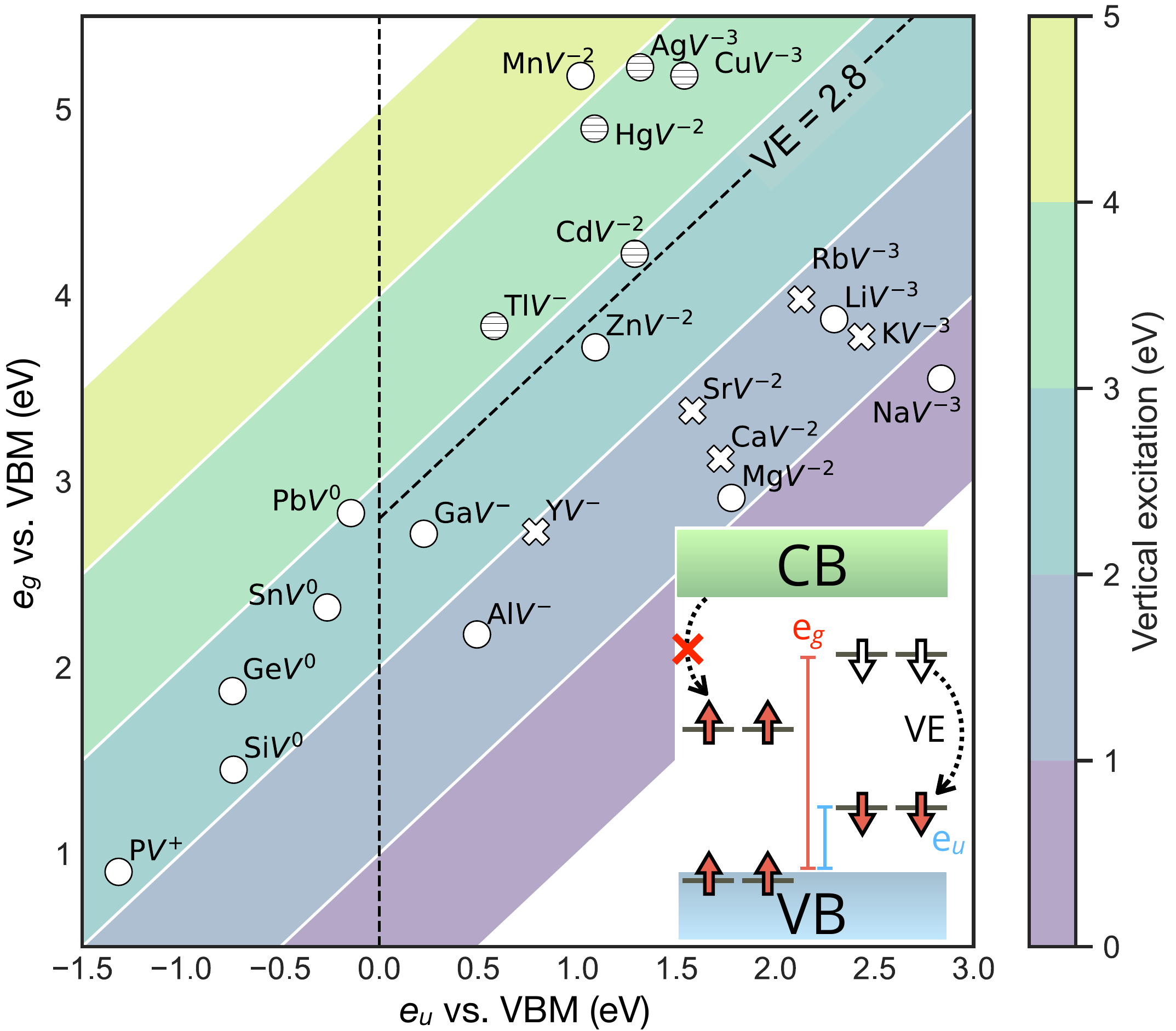}    
 	\caption{HSE$_0$ Kohn-Sham defect levels of $e_u$ and $e_g$ for all the defects in $D_{3d}$ with triplet ground state, similar to Zn$V^{-2}$. The inset depicts the design principle of the defect electronic structure. The empty circle, cross, and shaded circle stands for ``normal", ``inverted" and ``competing" electronic structures. See main text for details. $e_g$ and $e_u$ defect levels are measured with respect to the valence band maximum (VBM). Each HSE$_0$ Kohn-Sham diagrams are shown in Figure SX in SI.  
 	}
 	\label{eueg}
\end{figure}

Our high-throughput search has identified Zn$V^{2-}$ as a promising candidate. We now examine the substitution-vacancy defect database to gain insights into design principles for high-performance quantum defects based on their electronic structures. An overview of the symmetry for charged defects is summarized in Figure~\ref{Fig.periodic_table_trend}. In the periodic table representation, each element block is divided into eight sub-blocks that represent a distinct charge state. The colors indicate the symmetry point groups as shown by the label. The shaded blocks denote thermodynamically unstable charged defects. The general trend reveals that defects tend to exhibit D$_{3d}$ symmetry when the defect atoms belong to main group elements (alkali, alkaline-earth, and $p$-block elements) or late transition metals such as Zn or Cd. Conversely, most MV defects containing early transition metals distort toward lower symmetries. This distortion likely arises from significant contributions of transition metal $d$ orbitals to the in-gap states, creating complex bonding environments. Consequently, Jahn-Teller distortions occur as the $d$ orbitals become partially occupied, reducing the symmetry. In contrast, late transition metals feature deeper-lying $d$ orbitals with limited participation in defect orbitals, akin to main group elements, thus favoring higher-symmetry configurations. It is noted that the well-known Ni$V$ complex, a transition metal-vacancy defect exhibiting inversion symmetry, is also captured in our database \cite{Morris_arxiv_2024, Silkinis_RPB_2024}.

We have illustrated the chemical trends of defect complexes capable of adopting centrosymmetric configurations. Next, we discuss the criteria for defect levels to be appropriately positioned within the band gap. All defects with $D_{3d}$ symmetry and a triplet ground state are summarized in Figure~\ref{eueg}. These defects are mapped based on the energy differences between the $e_u$ and $e_g$ states relative to the valence band maximum (VBM). Here, we approximate the vertical excitation (VE) energy using the Kohn-Sham (KS) energy difference between the $e_u$ and $e_g$ states, as represented by the color bar. Two primary constraints emerge for appropriate defect level placement: firstly, the in-gap defect levels must be sufficiently high to avoid resonances with the band edges. Resonating $e_u$ states could lead to dark states due to optically forbidden transitions between the VBM ($a_{1g}$) and the defect levels. This situation is indicated by negative values of the $e_u$ state relative to the VBM, highlighted by a vertical dashed line. Such electronic structure is common for group IV vacancy complex defects. Secondly, defect levels should not be positioned too high, as this introduces competing excitations with the conduction band minimum (CBM) in the spin-majority channel. These scenarios are illustrated in the inset of Figure~\ref{eueg}. We observed that competing excitations (denoted by shaded markers) typically occur when vertical excitation energies exceed approximately 2.8 eV, except for MnV$^{2-}$. Most defects exhibiting competing excitations involve heavy, late transition metals. Thus, defects with the most favorable electronic structures tend to appear on the right-hand side of the map. Given that electronic and ionic relaxations are not explicitly included here, we anticipate sufficient emission energies for defects such as GaV$^{-}$ and Zn$V^{2-}$. Additionally, we note that in the spin-minority channel, $e_g$ states are not necessarily higher in energy than $e_u$ states—this primarily occurs in defect complexes involving alkali and alkaline-earth metals, as indicated by cross markers.

\section{Conclusion}

Through the construction and systematic screening of a substitution-vacancy complex defect database using an automated high-throughput computational workflow, we identified a series of candidates including Ga$V^-$, Ca$V^-$, Al$V^-$, and Zn$V^{2-}$ which we anticipate to emit brightly technologically relevant wavelength with vanishing Stark shift. We highlights the Zn$V^{2-}$ among the screened candidates. It as has a triplet ground state and appears to be an attractive quantum defect candidate in diamond. Zn$^{V2-}$ is readily accessible in terms of charge state and is thermodynamically favored compared to its constituent simple defects. Its $D_{3d}$ symmetry enhances robustness against fluctuating electric fields, significantly reducing spectral diffusion. The defect energy levels are well-separated from bulk bands, minimizing efficiency losses due to dark states, and positioned sufficiently low to avoid competing excitations across spin channels. Our comprehensive analysis of Zn$V^{2-}$ predicts its ZPL to be around 1.88 eV in the visible range, and with a DWF around 6.8\%. Additionally, our data-driven approach proposes design principles for achieving defects with inversion symmetry and optimized defect-level placement within the band gap. This study underscores the application of high-throughput computational methods for discovering quantum defects in wide bandgap semiconductors and encourages further experimental and theoretical research into Zn$^{V2-}$ for quantum technological applications.

\section{Methods}

The high-throughput defect computations were performed using the automatic workflows that are implemented in atomate software package \cite{Jain2013, Mathew2017,Ong2013}. All the first-principles calculations were performed using Vienna Ab-initio Simulation Package (VASP) \cite{G.Kresse-PRB96,G.Kresse-CMS96} with the projector augmented wave (PAW) method \cite{P.E.Blochl-PRB94}. For all the high-throughput computations, each charged defect is simulated in a supercell of 216 atoms with Perdew-Burke-Erzhenhoff (PBE) functionals\cite{J.Perdew-PRL96}. 520 eV cutoff energies were used for the plane-wave basis and only the $\Gamma$ point was used to sample the Brillouin zone. With a fixed volume, only the atomic positions are optimized until the ionic forces are smaller than 0.01 eV/\r{A} to simulate the dilute limit. Using the converged PBE wavefunction and charge density, we further performed single-shot HSE (HSE$_0$) improve the description of the electronic structure at a minimal cost \cite{Xiong2023, Heyd2003}. In total our database includes 2151 charge defects in diamond, with 539 substitution-vacancy complex defects and the others are simple point defects (substitutions and interstitials in tetrahedral and hexagonal). Each of these complex defects is initialized in low symmetry configuration similar to NV center and proceed to structure relaxation without symmetry constraint. This is employed to avoid the initial structure to be trapped in a high-symmetry local minimum.

For each screened candidate, we performed full HSE calculations in a 512-atom supercell with a reduced cutoff of 400 eV. The input generation and output analysis of the charge defects are performed using PyCDT \cite{Broberg2018}. The formation energy of each charged defect reads:

\begin{equation}
    E_\mathrm{form}[X^q] = E_\mathrm{tot}[X^q] - E_\mathrm{tot}^\mathrm{bulk} - \sum n_i \mu_i + q E_f + E_\mathrm{corr},
\label{eq:defect}
\end{equation}
where the formation energy is expressed as a function of the Fermi level $E_f$ for a given defect $X$ in the charge state $q$\cite{Zhang1991,Komsa2012}. $E_\mathrm{tot}[X^q]$, $E_\mathrm{tot}^\mathrm{bulk}$ corresponds to the total energies of the defect-containing supercell and bulk supercell. The energy needed to exchange atoms with thermodynamic reservoirs is represented by the number of atoms of species $n_i$ removed or added to create the defect, and the elemental chemical potential $\mu_i$ is employed in this work. Electronic chemical potential is given by the Fermi level $q E_f$. Finally, a correction term is applied to account for spurious image-charge Coulomb interactions due to finite supercell size, as well as potential-alignment corrections to restore the position of the bulk valence band maximum (VBM) in charged-defect calculations due to the presence of the compensating background charge density \cite{Freysoldt2011,Kumagai2014}. Here we used Kumagai correction throughout this work \cite{Kumagai2014}.

The transition dipole moment and radiative lifetime were evaluated using the single-particle wavefunction and further processed with PyVaspwfc code \cite{Zheng2018}. The radiative lifetime was approximated using Wigner-Weisskopf theory of fluorescence \cite{Gali2019,Alkauskas2016.PRB,Davidsson2020}: 

\begin{equation}
\frac{1}{\tau}=\frac{n_{r} (2 \pi)^3 \nu^{3}|{ \boldsymbol{\mu}}|^{2}}{3 \varepsilon_{0} h c^{3}},
\end{equation}
where $\tau$, $n_{r}$, ${\boldsymbol{\mu}}$, $\varepsilon_{0}$, $h$ and $c$ correspond to the radiative lifetime, refractive index of the host, transition dipole moment, vacuum permittivity, Planck constant, and the speed of light, respectively. $\nu$ is the transition frequency corresponding to the energy difference of the Kohn-Sham levels, and we used HSE$_0$ Kohn-Sham levels during the high-throughput calculations. The transition dipole moment is given as:

\begin{equation}
\boldsymbol{\mu}_{k}=\frac{\mathrm{i} \hbar}{\left(\epsilon_{\mathrm{f}, k}-\epsilon_{\mathrm{i}, k}\right) m}\left\langle\psi_{\mathrm{f}, k}|\mathbf{p}| \psi_{\mathrm{i}, k}\right\rangle,
\end{equation}
where $\epsilon_{\mathrm{i}, k}$ and $\epsilon_{\mathrm{f}, k}$ are the eigenvalues of the initial and final states, $m$ is the electron mass, $\psi_{\mathrm{i}}$ and $\psi_{\mathrm{f}}$ are the initial and final wavefunctions, and $\mathbf{p}$ is the momentum operator. The symmetries of the wavefunctions are assigned based on the analysis using Irvsp \cite{irvsp_2021}.

Photoluminescence lineshapes were simulated under Huang-Rhys approximation following the method proposed by Alkauskas \cite{Alkauskas2014, Razinkovas2021}, where the ground/excited states structures and phonon properties are simulated at HSE and GGA levels, respectively. All the phonon calculations were performed in a 511-atoms supercell. Phonon properties are processed using \texttt{Phonopy} and the compressive-sensing method implemented in \texttt{Pheasy}\cite{phononpy,pheasy,Zheng2024arxiv}. The ZPL positions were set to the values that were computed using HSE $\Delta$SCF method. PL lineshapes are plotted using \texttt{PyPhotonics}\cite{pyphotonics}.

\section{Acknowledgments}
\begin{acknowledgments}
This work was supported by the U.S. Department of Energy, Office of Science, Basic Energy Sciences 
in Quantum Information Science under Award Number DE-SC0022289. 
This research used resources of the National Energy Research Scientific Computing Center, 
a DOE Office of Science User Facility supported by the Office of Science of the U.S.\ Department of Energy 
under Contract No.\ DE-AC02-05CH11231 using NERSC award BES-ERCAP0020966. 
\end{acknowledgments}

\bibliography{main}
\end{document}